\documentclass[aps,prl,twocolumn,showpacs,preprintnumbers,amsmath,amssymb,superscriptaddress]{revtex4}%

\usepackage{graphicx}%
\usepackage{dcolumn}
\usepackage{amsmath}
\usepackage{color}
\usepackage{multirow}

\begin{document}

%\preprint{PREPRINT (\today)}

\title{Low-temperature magnetic fluctuations in the Kondo insulator SmB$_6$}

\author{P.~K.~Biswas}
\email[Corresponding author: ]{pabitra.biswas@psi.ch}
\affiliation{Laboratory for Muon Spin Spectroscopy, Paul Scherrer Institut, CH-5232 Villigen PSI, Switzerland}

\author{Z.~Salman}
\affiliation{Laboratory for Muon Spin Spectroscopy, Paul Scherrer Institut, CH-5232 Villigen PSI, Switzerland}

\author{T.~Neupert}
\affiliation{Princeton Center for Theoretical Science, Princeton University, Princeton, New Jersey 08544, USA}

\author{E.~Morenzoni}
\affiliation{Laboratory for Muon Spin Spectroscopy, Paul Scherrer Institut, CH-5232 Villigen PSI, Switzerland}

\author{E.~Pomjakushina}
\affiliation{Laboratory for Developments and Methods, Paul Scherrer Institut, CH-5232 Villigen PSI, Switzerland}

\author{F.~von~Rohr}
\affiliation{Physik-Institut der Universit$\ddot{a}$t Z$\ddot{u}$rich, Winterthurerstrasse 190, CH-8057 Z$\ddot{u}$rich, Switzerland}

\author{K.~Conder}
\affiliation{Laboratory for Developments and Methods, Paul Scherrer Institut, CH-5232 Villigen PSI, Switzerland}

\author{G.~Balakrishnan}
\affiliation{Department of Physics, University of Warwick, Coventry, CV4 7AL, UK}

\author{M.~Ciomaga~Hatnean}
\affiliation{Department of Physics, University of Warwick, Coventry, CV4 7AL, UK}

\author{M.~R.~Lees}
\affiliation{Department of Physics, University of Warwick, Coventry, CV4 7AL, UK}

\author{D.~McK.~Paul}
\affiliation{Department of Physics, University of Warwick, Coventry, CV4 7AL, UK}

\author{A.~Schilling }
\affiliation{Physik-Institut der Universit$\ddot{a}$t Z$\ddot{u}$rich, Winterthurerstrasse 190, CH-8057 Z$\ddot{u}$rich, Switzerland}

\author{C.~Baines}
\affiliation{Laboratory for Muon Spin Spectroscopy, Paul Scherrer Institut, CH-5232 Villigen PSI, Switzerland}

\author{H.~Luetkens}
\affiliation{Laboratory for Muon Spin Spectroscopy, Paul Scherrer Institut, CH-5232 Villigen PSI, Switzerland}

\author{R.~Khasanov}
\affiliation{Laboratory for Muon Spin Spectroscopy, Paul Scherrer Institut, CH-5232 Villigen PSI, Switzerland}

\author{A.~Amato}
\affiliation{Laboratory for Muon Spin Spectroscopy, Paul Scherrer Institut, CH-5232 Villigen PSI, Switzerland}

\begin{abstract}
We present the results of a systematic investigation of the magnetic properties of the 3D Kondo topological insulator SmB$_6$ using magnetization and muon spin relaxation/rotation ($\mu$SR) measurements. The $\mu$SR measurements exhibit magnetic field fluctuations in SmB$_6$ below $\sim$~15~K due to electronic moments present in the system. However, no evidence for magnetic ordering is found down to 19~mK. The observed magnetism in SmB$_6$ is homogeneous in nature throughout the full volume of the sample. Bulk magnetization measurements on the same sample show consistent behavior. The agreement between $\mu$SR, magnetization and NMR results strongly indicate the appearance of intrinsic bulk magnetic in-gap states associated with fluctuating magnetic fields in SmB$_6$ at low temperature.
\end{abstract}

\maketitle

%\section{Introduction}

Kondo insulators are mostly realized in strongly correlated rare-earth material systems. At high temperature, these materials behave as highly correlated metals, while at low temperature they are simply band insulators due to the formation of an energy gap at the Fermi level \cite{Aeppli,Riseborough,Coleman}. The opening of a gap at low temperature is attributed to hybridization between the localized $f$ electrons (mostly from unfilled 4$f$-shells of the rare earth atoms) and  the conduction electrons. SmB$_6$, a mixed valence heavy fermion compound, more frequently referred to as Kondo insulator (even though Sm has non-integer chemical valence close to 2.5), has been very well known for many years due to its exotic low temperature transport properties. In this material, as the temperature is reduced, its resistivity increases exponentially as expected for a normal insulator. However, as the temperature is reduced further below 4~K, the resistivity saturates at a finite value (a few $\Omega$.cm)~\cite{Menth}. This behavior was attributed to certain ``in-gap'' states~\cite{Cooley}, whose true nature was revealed only recently, when SmB$_6$ was predicted theoretically to be a 3D topological insulator. As such, it features topologically protected metallic surface states at low temperature \cite{Dzero1,Soluyanov,Yu,Dzero2}, which lie in the bulk gap. Several ARPES measurements conducted on SmB$_6$ reveal a Kondo gap of a few meV in the bulk and also identify the low-lying in-gap states close to the Fermi level~\cite{Neupane,Xu,Miyazaki,Jiang,Denlinger}. These in-gap states are found to disappear as the temperature is raised above $\sim$~15~K~\cite{Neupane}. Although, other ARPES results suggest that the transition is very broad and that the in-gap states disappear completely at a much higher temperature~\cite{Xu}. A very recent ARPES study has further suggested that the in-gap states gradually transform from 2D to 3D nature with increasing temperature~\cite{Denlinger2,Denlinger3}. These insights are complemented by surface-related transport measurements which also suggest that the surface conductivity can be ascribed to topologically protected surface states~\cite{Wolgast,Kim,Zhang,Kim2}. 

Besides this intriguing charge response, SmB$_6$ also shows peculiar magnetic properties. NMR measurements have shown an enhanced spin lattice relaxation in high applied magnetic fields \cite{Caldwell07PRB,Takigawa81JPSJ,Pena81SSC}, which could be attributed to a contribution from magnetic in-gap states to the nuclear relaxation below $\sim 10$ K~\cite{Caldwell07PRB}. These magnetic in-gap states are not to be confused with the topological surface in-gap states mentioned above. They are true bulk excitations. The possibilities that they arise as bound states at B$_6$ vacancies or other impurities were ruled out~\cite{Caldwell07PRB}. It will be crucial to clarify the nature of these magnetic in-gap states, in order to obtain a complete understanding of the low-energy physics of the topological Kondo insulator SmB$_6$. In addition, no detailed experimental work was yet performed to search for possible intrinsic Sm moments in the absence of applied magnetic field, in particular at low temperature close to the transport anomaly regime.

Here, we report the results of muon-spin relaxation ($\mu$SR) studies on two different samples of SmB$_6$. In both samples, we observe a clear signature of fluctuating local magnetic fields appearing below $\sim$~15~K, which corresponds to the regime in which the magnetic in-gap states have been conjectured. We do not see any ordering of the fluctuating magnetic fields down to 19~mK. Magnetization measurements show consistent magnetic behavior. Our results strongly suggest that the appearance of bulk magnetic in-gap states is intrinsic to SmB$_6$.

%\section{Experimental details}\label{sec:experimental}

Muon spin rotation/relaxation ($\mu$SR) measurements were performed using the DOLLY spectrometer at PSI, Switzerland. Low temperature $\mu$SR data were collected on the LTF spectrometer down to 19~mK. Measurements on two different single crystal samples are reported here; sample A was a large single crystal and grown using the floating-zone method, while sample B consisted of small crystals which were grown using Al-flux. In a bulk-$\mu$SR experiment, $100\%$ spin polarized (along the direction of the muon beam, $z$) positive muons are implanted in the bulk of the sample. The implanted muons decay (lifetime, $\tau_\mu=2.2$~$\mu$s) into positrons emitted preferentially in the direction of the muon-spin direction at the time of decay. In zero field (ZF) and longitudinal field (LF) $\mu$SR measurements we measure the asymmetry of the muon decay along the $z$-axis as a function of time, $A(t)$, by detecting and time-stamping the decay positrons using positron detectors, placed in forward (\textit{F}) and backward (\textit{B}) direction with respect to the initial muon spin direction. The positron counts in the $F$ and $B$ detectors, $N_{F,B}\left(t\right)$ have the following functional form:
\begin{equation}
N_{F,B}\left(t\right)=N_{F,B}\left(0\right)\exp\left(-\frac{t}{\tau_\mu}\right)\left[1\pm A\left(t\right)\right].
\end{equation}
$A(t)$ is determined by using the following equation: 
\begin{equation}
A\left(t\right)=\frac{N_{F}\left(t\right)-\alpha N_{B}\left(t\right)}{N_{F}\left(t\right)+\alpha N_{B}\left(t\right)},
\end{equation}
where $\alpha$ is a parameter taking into account differences in geometry and efficiency of the positron detectors; it was determined in a separate calibration measurement. $A(t)$ is determined by the static and dynamic properties of the local fields probed by the muons. For more details about this technique and its use, see Ref.~\cite{Dalmas} and references therein. All the $\mu$SR data were analyzed using the MUSRFIT package~\cite{Suter}.

%%\section{Results and discussion}\label{sec:results}

\begin{figure}[htb]
\includegraphics[width=1.0\linewidth]{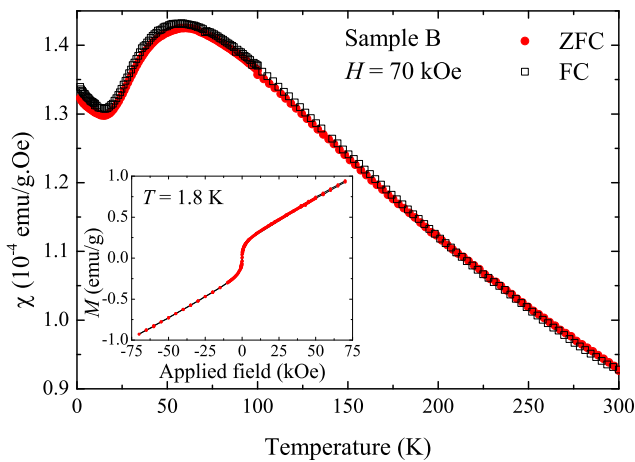}
% \vspace{-1cm}
\caption{(Color online) Temperature dependence of the magnetic susceptibility of SmB$_6$, measured under ZFC and FC conditions in an applied field of 70~kOe. Inset shows the full $M(H)$ loop of SmB$_6$, measured at 1.8~K.}
 \label{fig:SmB6_MvTvH}
\end{figure}

The magnetization measurements were performed using a Quantum Design SQUID magnetometer. Figure~\ref{fig:SmB6_MvTvH} shows the magnetic susceptibility $\chi$ as a function of temperature. The data were collected under zero field cooled (ZFC) and field cooled (FC) conditions in an applied field of 70~kOe. There is no opening between the ZFC and FC data, indicating lack of irreversible or history dependent magnetism in SmB$_6$. A dome shaped magnetization curve is observed in the $\chi(T)$ data at $\sim$~55 K. The magnetic nature seems to change at temperatures below $\sim$~15~K, where a pronounced upturn is observed which does not show the typical behavior of paramagnetic impurities. The inset in Fig.~\ref{fig:SmB6_MvTvH} shows the full $M(H)$ loop of SmB$_6$, collected at 1.8~K. There is no observable hysteresis in the $M(H)$ loop, which is consistent with the susceptibility curves. However, a clear deviation from a linear paramagnetic response is observed at low fields (see the inset of Fig.~\ref{fig:SmB6_MvTvH}). Note that, extensive studies of the magnetic and structural properties of sample A found no evidence for defects or impurities~\cite{CiomagaHatnean13SR}.
\begin{figure}[htb]
\includegraphics[width=1.0\linewidth]{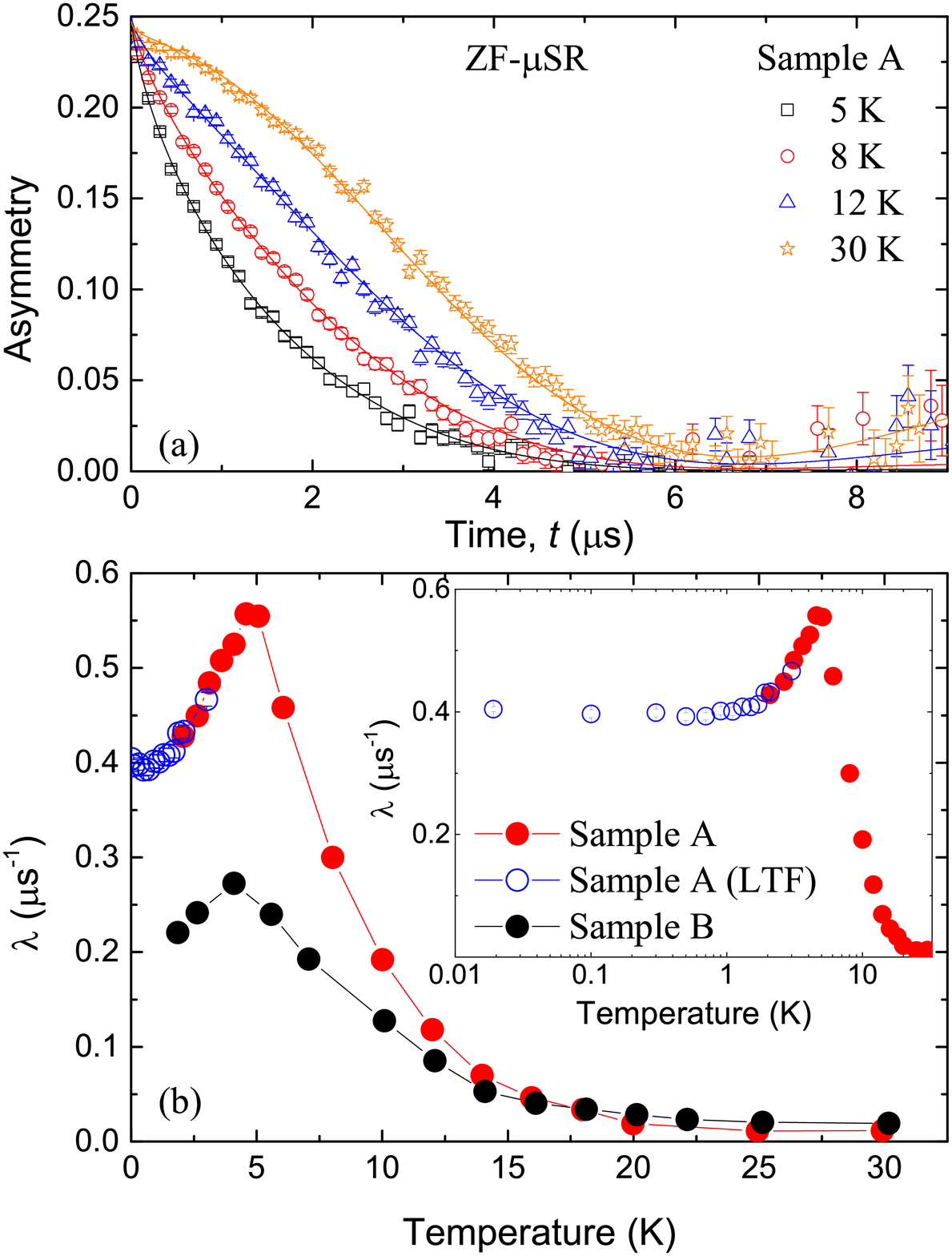}
% \vspace{-1cm}
\caption{(Color online) (a) ZF-$\mu$SR asymmetry signals of SmB$_6$, collected at different temperatures for Sample A. The solid lines are fits to the data using Eq.~\ref{eq:KT_ZFequation}. (b) Temperature dependence of the muon spin relaxation rate $\lambda$ due to the presence of electronic moments in SmB$_6$. Inset shows the $\lambda(T)$ of sample A in logarithmic scale.}
 \label{fig:SmB6_ZF}
\end{figure}

ZF-$\mu$SR is a very sensitive probe of the intrinsic magnetism in a material. Figure~\ref{fig:SmB6_ZF}(a) shows typical ZF-$\mu$SR asymmetries, collected at different temperatures for Sample A. The asymmetry at 30~K displays a Gaussian-like muon spin relaxation which is caused only by randomly oriented nuclear dipole moments~\cite{Dalmas}. However, with decreasing temperature, clear changes in the shapes of the asymmetry are observable. These are indicative of the appearance of additional dilute local magnetic fields, most probably due to electronic magnetic moments in SmB$_6$. The ZF asymmetry as a function of temperature was found to fit best to a Kubo-Toyabe relaxation function multiplied by a stretched exponential decay function of the following form:
\begin{equation}
A(t)= A_0\left\{\frac{1}{3}+\frac{2}{3}\left(1-a^2t^2\right){\rm exp}\left(-\frac{a^2t^2}{2}\right)\right\}\exp(-\lambda t)^\beta,
 \label{eq:KT_ZFequation}
\end{equation}
where $A_0$ is the initial asymmetry, $\beta$ is the stretch parameter, and $a$ and $\lambda$ are the muon spin relaxation rates due to the presence of static nuclear moments and electronic moments, respectively. Global fits yield good agreement to the measured temperature dependence of all ZF data with common parameters $a=0.253(1)$~$\mu$s$^{-1}$, and $\beta=0.74(1)$ for sample A and $a=0.253(1)$~$\mu$s$^{-1}$, and $\beta=0.72(1)$ for sample B. Figure~\ref{fig:SmB6_ZF}(b) shows the fitted parameter $\lambda$ as a function of temperature. We observe a large increase in $\lambda$ below $\sim$~15~K for both samples, indicating the onset of a magnetic signal in SmB$_6$ below this temperature. We find a pronounced peak at $\sim$~5~K. We attribute this peak to changes in the dynamics of the local magnetic fields appearing below $\sim$~15~K. For sample A, we collected ZF data down to 19~mK. To show the low temperature behavior more clearly, $\lambda(T)$ is plotted in logarithmic scale in the inset of Fig.~\ref{fig:SmB6_ZF}~(b). The relaxation rate decreases continuously below 5~K and then saturates and becomes temperature independent below $\sim$~2~K. Therefore, it is likely that the fluctuations of the local magnetic moments in SmB$_6$ are driven by non-thermal processes below this temperature. Furthermore, we do not see any indication of magnetic ordering down to 19~mK. Both samples show a similar qualitative temperature dependence, although $\lambda$ is smaller in sample B. The origin of this difference is not clear. This could be due to contamination with non-relaxing background signal in the small sample which will effectively lower the relaxation rate. Another reason may be the two completely different growth processes of the single crystals, lead to slightly different microscopic properties. It is noteworthy that Eq.~\ref{eq:KT_ZFequation} accounts for the full signal, indicating a homogeneous Sm moments in the full volume of the sample below $\sim$~15~K. Therefore, the observed behavior cannot be attributed to an impurity phase in the samples.
\begin{figure}[htb]
\includegraphics[width=1.0\linewidth]{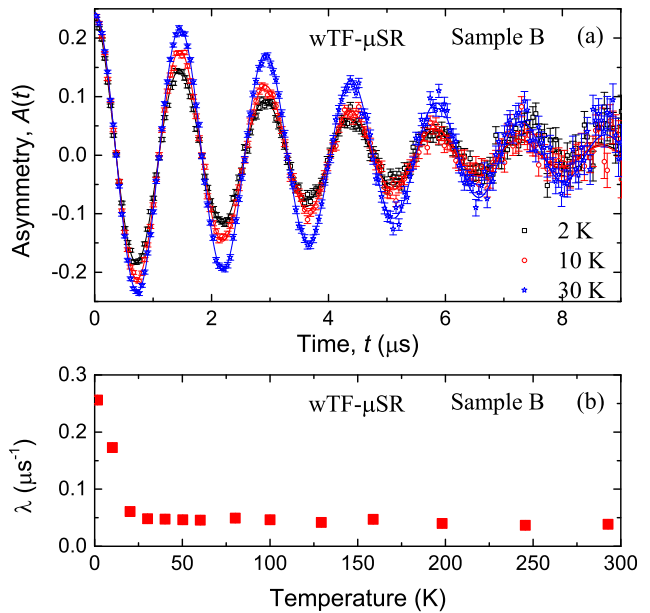}
% \vspace{-1cm}
\caption{(Color online) (a) The wTF-$\mu$SR asymmetries, collected at different temperatures for Sample B in a wTF of 50~Oe. (b) The temperature dependence of $\lambda$ of SmB$_6$, collected from the wTF-$\mu$SR data.}
 \label{fig:SmB6_wTF}
\end{figure}

This is confirmed by $\mu$SR experiments under a weak transverse-field (wTF) of 50 Oe. Figure 3 (a) shows the wTF-$\mu$SR asymmetries, collected at different temperatures for sample B. The temperature independence of the signal amplitude $A_0$ rules out the possibility of a magnetic impurity phase in the sample. wTF-$\mu$SR data also shows similar decay of the precession signals as seen in the ZF-$\mu$SR relaxation signals. The fits to the data were made using
\begin{equation}
A(t)= A_0\cos\left(\omega t+\phi\right){\rm exp}\left(-\frac{\sigma^2t^2}{2}\right)\exp(-\lambda t)^\beta,
 \label{eq:KT_wTFequation}
\end{equation}
where $\omega$ is the precession frequency, $\phi$ is the initial polarization phase, and $\lambda$ is the relaxation rates of the muon precession signal due to electronic moments. Again in these fits, all the parameters were taken temperature independent except for $\lambda$. For consistency, we assumed a fixed value of $\beta=0.72(1)$ as obtained from ZF-$\mu$SR. A global fit to all wTF-$\mu$SR data yields $\sigma=0.207(42)$~$\mu$s$^{-1}$. Figure~\ref{fig:SmB6_wTF}(b) shows the temperature dependence of $\lambda$, where $\lambda(T)$ shows a sharp increase below ~15 K in agreement with the ZF results.

\begin{figure}[htb]
\includegraphics[width=1.0\linewidth]{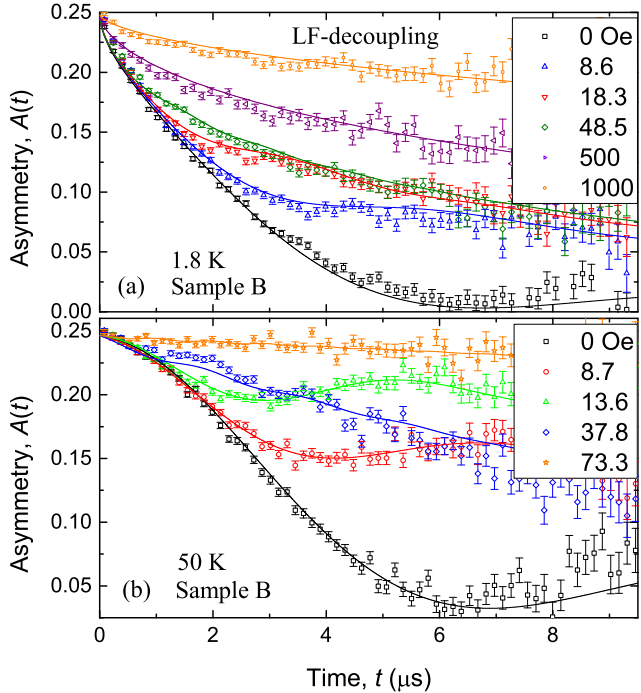}
% \vspace{-1cm}
\caption{(Color online) Asymmetry signals under different applied longitudinal fields, collected at (a) 1.8~K and (b) 50~K. The solid lines are the fits to the data using Eq.~\ref{eq:KT_LFequation}.}
 \label{fig:SmB6_LFdecoup}
\end{figure}

To better understand the nature of the local magnetic fields appearing at low temperature, i.e. whether these are static or dynamic, we have performed decoupling measurements in LF. Figures~\ref{fig:SmB6_LFdecoup}(a) and \ref{fig:SmB6_LFdecoup}(b) show $A(t)$ measured at 1.8~K and 50~K, respectively, with different magnetic fields applied along the initial direction of the polarization. Note that at high temperature a small field of $\sim$~70~Oe completely decouples the muon spin, whereas at 1.8~K signal relaxation is still observed at 1~kOe. This shows that at high temperature the local field sensed by the muons are simply the static nuclear fields, which can be easily decoupled, while at low temperatures low frequency fluctuating fields are present, consistent with the appearance  of magnetic in-gap states in SmB$_6$ below $\sim$~15~K. The fits to the LF-$\mu$SR signals were made with a function of the following form:
\begin{equation}
A(t)= A_0 P_{GKT}(B,\Delta,t)\exp(-\lambda t)^\beta,
 \label{eq:KT_LFequation}
\end{equation}
where $P_{GKT}(B,\Delta,t)$ is the Gaussian Kubo-Toyabe depolarization function in the applied field $B$, with static fields distribution width $\Delta$, and $\lambda$ is the relaxation rate due to the dynamic fields. The solid lines in Figure~\ref{fig:SmB6_LFdecoup}(a) are the resulting fits , where we assume that $A_0$, $\Delta$, and $\beta$ are independent of field. For consistency, we have fixed $\beta=0.72(1)$ to the value obtained from the ZF-$\mu$SR data fit. A global fit to all the LF-$\mu$SR data yield $\Delta=0.265(3)$~$\mu$s$^{-1}$ at 1.8~K and $\Delta=0.250(2)$~$\mu$s$^{-1}$ at 50~K. The value of $\Delta$ for the nuclear contribution, is consistent with the values we extracted from ZF- and wTF-$\mu$SR data fits, and shows that there is {\em no temperature dependence} in the static fields present in SmB$_6$.

\begin{figure}[htb]
\includegraphics[width=1.0\linewidth]{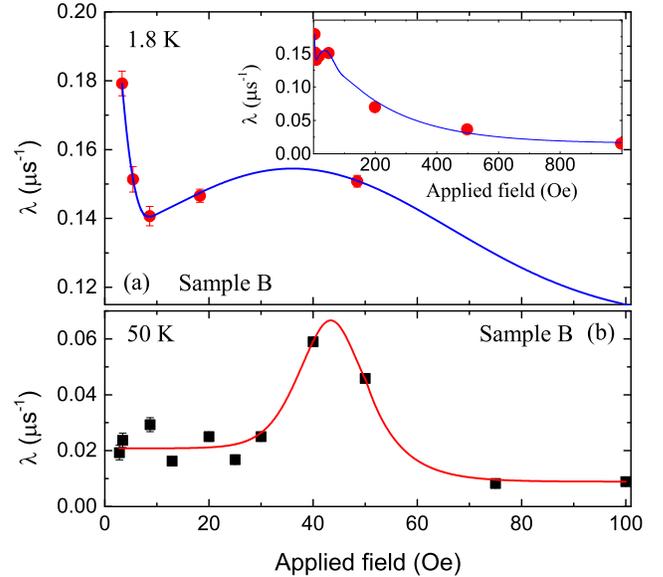}
% \vspace{-1cm}
\caption{(Color online) Field dependence of $\lambda$ for (a) 1.8~K and (b) 50~K, showing the LCR peaks at around 40~Oe for both temperatures. The solid lines are a guide to the eye. The inset in Fig.~~\ref{fig:SmB6_LFdecoup_lambda}~(a) shows the field dependence of $\lambda$ up to 1000~Oe.}
 \label{fig:SmB6_LFdecoup_lambda}
\end{figure}

Figures~\ref{fig:SmB6_LFdecoup_lambda}(a) and \ref{fig:SmB6_LFdecoup_lambda}(b) show the variation of $\lambda$ as a function of applied field at 1.8~K and 50~K, respectively. The inset in Fig.~\ref{fig:SmB6_LFdecoup_lambda}~(a) shows the field dependence of $\lambda$ up to 1000~Oe. We first note that the values of $\lambda$ are much higher at 1.8~K than 50~K, a clear indication that the dynamic contribution to the muon spin relaxation at low temperatures is dramatically larger than at high temperatures. Moreover, while no field dependence is observed in $\lambda$ in the low field range at 50~K, an upturn is observed at 1.8~K. This is again evidence of the presence of slowly fluctuating local moments at low temperatures. It is noteworthy that similar low temperature dynamic magnetic correlations have also been observed in the past in another Kondo insulator YbB$_{12}$~\cite{Yaouanc}. For this system, it was speculated that the low-temperature upturn in the susceptibility may be due to impurities (oxidized Yb) on the grain boundaries or the surface. In the case of SmB$_6$, this cannot be the reason as we have used very clean single crystal samples and found no indication of impurity phases. Furthermore, persistent spin dynamics at low temperatures have been observed in $\mu$SR experiments on many \textit{f}-electron systems~\cite{Bertin,Yaouanc2,Yaouanc3}. However, frustration is unlikely to play any role in SmB$_6$. The peak observed around 40~Oe for both temperatures is due to level crossing relaxation (LCR). A LCR is observed when the energy level splitting in the host systems matches that of the probe, such that energy exchange between the two is possible and an enhanced loss/relaxation of polarization is observed~\cite{Kreitzman}. The observed LCR in SmB$_6$ is most probably due to a matching condition between B nuclei and the implanted muons. A similar LCR peak has been observed in CaB$_6$ single crystals at around 70~Oe below $\sim$~100~K using the $\mu$SR technique~\cite{Gygax}. 

In conclusion, we observe a signature of slowly fluctuating magnetic fields appearing below $\sim$~15~K, probably due to fluctuating intrinsic electronic magnetic moments, which do not order magnetically down to 19~mK. Bulk magnetization measurements on the same samples also show a magnetic ``anomaly" below $\sim$~15~K and a clear deviation from a simple para/diamagnetic behavior at low temperatures. The $\mu$SR data further indicate that the magnetic properties of SmB$_6$ are homogeneous in the full volume of the sample. Finally, we conclude that the magnetic fields appearing at low temperature are dynamic in nature, which may still preserve the time reversal symmetry in this system. The presence of such magnetic fluctuation in a topological insulator cannot be justified by any current theory and hence demand further theoretical investigation. It is widely known that a topological surface state is protected by time-reversal-symmetry which is not valid in the presence of magnetic ordering. The measurements reported here provide information regarding the bulk of SmB$_6$, which may be different near the surface. Therefore, further studies using near-surface sensitive techniques, such as low energy $\mu$SR~\cite{Morenzoni94PRL,Prokscha08NIMA}, are necessary to detect any changes in the magnetic behavior near the surface of this system. These bulk-$\mu$SR studies of SmB$_6$ will provide an excellent reference for such future studies.

This work was performed at the Swiss Muon Source (S$\mu$S), Paul Scherrer Institute (PSI, Switzerland). The work was supported, in part, by the EPSRC, United Kingdom grant no. EP/I007210/1.


\begin{thebibliography}{99}
%
\bibitem{Aeppli} G.~Aeppli, and Z.~Fisk, Comm.~Condens.~Matter~Phys. {\bf 16}, 155 (1992).
%
\bibitem{Riseborough} P.~S.~Riseborough, Adv.~Phys. {\bf 49}, 257 (2000).
%
\bibitem{Coleman} P.~Coleman, in \textit{Handbook~of~Magnetism~and~Advanced~Magnetic~Materials} (Wiley, New York 2007), Vol. 1.
%
\bibitem{Menth} A.~Menth, E.~Buehler, and T. H.~Geballe, Phys.~Rev.~Lett. {\bf 22}, 295 (1969).
%
\bibitem{Cooley} J. C.~Cooley, M. C.~Aronson, Z.~Fisk, and P. C.~Canfield, Phys.~Rev.~Lett. {\bf 74}, 1629 (1995).
%
\bibitem{Dzero1} M.~Dzero, K.~Sun, V.~Galitski, and P.~Coleman, Phys.~Rev.~Lett. {\bf 104}, 106408 (2010).
%
\bibitem{Soluyanov} A. A.~Soluyanov, and D.~Vanderbilt, Phys.~Rev.~B {\bf 83}, 035108 (2011).
%
\bibitem{Yu} R.~Yu, X.-L.~Qi, A.~Bernevig, Z.~Fang, and X.~Dai, Phys.~Rev.~B {\bf 84}, 075119 (2011).
%
\bibitem{Dzero2} M.~Dzero, K.~Sun, P.~Coleman, and V.~Galitski, Phys.~Rev.~B {\bf 85}, 045130 (2012).
%
\bibitem{Xu} N.~Xu, X.~Shi, P. K.~Biswas, C. E.~Matt, R. S.~Dhaka, Y.~Huang, N. C.~Plumb, M.~Radovic, J. H.~Dil, E.~Pomjakushina, A.~Amato, Z.~Salman, D. M.~Paul, J.~Mesot, H.~Ding, and M.~Shi, Phys.~Rev.~B {\bf 88}, 121102(R) (2013).
%
\bibitem{Neupane} M.~Neupane, N.~Alidoust, S.-Y.~Xu, T.~Kondo, D.-J.~Kim, C.~Liu, I.~Belopolski, T.-R.~Chang, H.-T.~Jeng, T.~Durakiewicz, L.~Balicas, H.~Lin, A.~Bansil, S.~Shin, Z.~Fisk, and M. Z.~Hasan, arXiv:1306.4634.
%
\bibitem{Miyazaki} H.~Miyazaki, T.~Hajiri, T.~Ito, S.~Kunii, and S.-I.~Kimura, Phys.~Rev.~B {\bf 86}, 075105 (2012).
%
\bibitem{Jiang} J.~Jiang, S.~Li, T.~Zhang, Z.~Sun, F.~Chen, Z. R.~Ye, M.~Xu, Q. Q.~Ge, S. Y.~Tan, X. H.~Niu, M.~Xia, B. P.~Xie, Y. F.~Li, X. H.~Chen, H. H.~Wen, and D. L.~Feng, arXiv:1306.5664.
%
\bibitem{Denlinger} J. D.~Denlinger, G.-H.~Gweon, J. W.~Allen, C. G.~Olson, Y.~Dalichaouch, B.-W.~Lee, M. B.~Maple, Z.~Fisk, P. C.~Canfield, and P. E.~Armstrong, Physica~B {\bf 281-282}, 716 (2000).
%
\bibitem{Denlinger2} J. D. Denlinger, J. W. Allen, J.-S. Kang, K. Sun, B.-I. Min, D.-J. Kim, and Z.~Fisk, arXiv:1312.6636 (2013).
%
\bibitem{Denlinger3} J. D. Denlinger, J. W. Allen, J.-S. Kang, K. Sun, J.-W. Kim, J. H. Shim, B.-I. Min, D.-J. Kim, and Z.~Fisk, arXiv:1312.6637 (2013).
%
\bibitem{Wolgast} S.~Wolgast, Y. S.~Eo, C.~Kurdak, K.~Sun, J. W.~Allen, D.-J.~Kim, and Z.~Fisk, arXiv:1211.5104 (2013).
%
\bibitem{Kim} D.-J.~Kim, S.~Thomas, T.~Grant, J.~Botimer, Z.~Fisk, and J.~Xia, Sci. Rep. \textbf{3}, 3150 (2013).
%
\bibitem{Zhang} X.~Zhang, N. P.~Butch, P.~Syers, S.~Ziemak, R. L.~Greene, and J.~Paglione, Phys.~Rev.~X  {\bf 3}, 011011 (2013).
%
\bibitem{Kim2} D.-J.~Kim, J.~Xia, and Z.~Fisk, arXiv:1307.0448 (2013).
%
\bibitem{Caldwell07PRB} T.~Caldwell, A.~P.~Reyes, W.~G.~Moulton, P.~L.~Kuhns, M.~J.~R.~Hoch, P.~Schlottmann, and Z.~Fisk, Phys. Rev. B {\bf 75}, 075106 (2007).
%
\bibitem{Takigawa81JPSJ} M.~Takigawa, H.~Yasuoka, Y.~Kitaoka, T.~Tanaka, H.~Nozaki, and Y.~Ishizawa, J. Phys. Soc. Jpn. {\bf 50}, 2525 (1981).
%
\bibitem{Pena81SSC} O.~Pe\~na, M.~Lysak, D. E.~MacLaughlin, and Z.~Fisk, Solid State Commun. {\bf 40}, 539 (1981).
%
\bibitem{Dalmas} P.~Dalmas~de~R\'{e}otier, and A.~Yaouanc, J.~Phys.:~Condens.~Matter {\bf 9}, 9113 (1997).
%
\bibitem{Suter} A.~Suter, and B.~M.~Wojek, Physics~Procedia {\bf 30}, 69 (2012).
%
\bibitem{CiomagaHatnean13SR} M.~Ciomaga Hatnean, M. R. Lees, D. M. Paul, and G. Balakrishnan, Sci. Rep. {\bf 3}, (2013).
%
%\bibitem{Sonier} J.E.~Sonier, J.H.~Brewer, and R.F.~Kiefl, Rev.~Mod.~Phys. {\bf 72}, 769 (2000).
%
%\bibitem{Brandt} E.H.~Brandt, Phys.~Rev.~B {\bf 37}, 2349 (1988).
%
\bibitem{Yaouanc} A.~Yaouanc, P.~Dalmas~de~R\'{e}otier, P. Bonville, G. Lebras, P. C. M. Gubbens, A. M. Mulders, and S. Kunii, Europhys. Lett. {\bf 47}, 247 (1999).
%
\bibitem{Bertin} E. Bertin, P. Bonville, J.-P. Bouchaud, J. A. Hodges, J. P. Sanchez, and P. Vulliet, Eur. Phys. J. B {\bf 27}, 347 (2002).
%
\bibitem{Yaouanc2} A. Yaouanc, P.~Dalmas~de~R\'{e}otier, V. Glazkov, C. Marin, P. Bonville, J. A. Hodges, P. C. M. Gubbens, S. Sakarya, and C. Baines, Phys. Rev. Lett. {\bf 95}, 047203 (2005).
%
\bibitem{Yaouanc3} A. Yaouanc, P. Dalmas de Reotier, Y. Chapuis, C. Marin, S. Vanishri, D. Aoki, B. Fak, L. P. Regnault, C. Buisson, A. Amato \textit{et al.}, Phys. Rev. B {\bf 84}, 184403 (2011).
%
\bibitem{Kreitzman} S.~R.~Kreitzman, Hyperfine~Interact. {\bf 31}, 13 (1986).
%
\bibitem{Gygax} F.~N.~Gygax, and A.~Schenck, J. Alloy Compd. {\bf 404}, 360 (2005).
%
\bibitem{Morenzoni94PRL}
E.~Morenzoni, F.~Kottmann, D.~Maden, B. Matthias, M. Meyberg, T. Prokscha, T. Wutzke, and U. Zimmermann, Phys. Rev. Lett.{\bf 72}, 2793 (1994).

\bibitem{Prokscha08NIMA}
T. Prokscha, E. Morenzoni, K. Deiters, F. Foroughi, D.~George, R. Kobler, A. Suter, and V. Vrankovic, Nucl. Instr. and Meth. A {\bf 595}, 317 (2008).

\end{thebibliography}
\end{document}